\documentclass[preprint,12pt]{elsarticle}

\usepackage{amssymb}
\usepackage{amsmath}
\usepackage{amsthm}

\usepackage{subcaption}

\newtheorem{prop}{Proposition}

\journal{Physics Letters A}

\begin{document}

\begin{frontmatter}



\title{A symbiotic SIR process} 


\author[label1]{Gerardo Palafox-Castillo} 
\author[label2]{Ericka Fabiola V\'azquez-Alcal\'a}
\author[label1,label2]{Arturo Berrones-Santos}
\affiliation[label1]{organization={Facultad de Ciencias F\'isico Matem\'aticas, Universidad Aut\'onoma de Nuevo Le\'on},
            addressline={Pedro de Alba S/n}, 
            city={San Nicol\'as de los Garza},
            postcode={66455}, 
            state={Nuevo Le\'on},
            country={M\'exico}}
\affiliation[label2]{organization={Facultad de Ingenier\'ia Mec\'anica y El\'ectrica, Universidad Aut\'onoma de Nuevo Le\'on},
            addressline={Pedro de Alba S/n}, 
            city={San Nicol\'as de los Garza},
            postcode={66455}, 
            state={Nuevo Le\'on},
            country={M\'exico}}

\begin{abstract}
We study a symmetric two-disease SIR co-infection model on networks in which co-infected individuals recover at a rate distinct from that of single infections. The model explicitly represents all co-infection states and features absorbing recovered compartments for both diseases. Within a mean-field network approximation, we derive the basic reproduction number of the coupled system and show that invasion thresholds coincide with those of two independent SIR processes. Exploiting an exchange symmetry in the equal-transmission regime, we reduce the dynamics to a lower-dimensional invariant subsystem and analyze the impact of the co-infection recovery rate. We prove that slower recovery of co-infected individuals monotonically increases the co-infection burden and yields a lower bound on epidemic duration that grows as the co-infection recovery rate decreases. Numerical simulations further indicate that reduced co-infection recovery can increase the epidemic peak, an effect supported by a sensitivity-equation analysis. Together, these results highlight how co-infection-specific recovery dynamics can substantially alter transient epidemic behavior, even in the absence of endemic equilibria.
\end{abstract}

\end{frontmatter}

\section{Introduction}
In a seminal work, \citet{de2012symbiotic} introduced a variation of the traditional contact process in which two species coexist and reproduce independently, but experience reduced mortality when sharing a site. Initially formulated on lattices, the model has since been extended to complex networks \citep{de2019dynamical,costa2022heterogeneous} and to higher-dimensional lattices \citep{sampaio2018symbiotic,Durrett2020-zs}. Formally, the symbiotic two-species contact process (2SCP) is defined as follows \citep{costa2022heterogeneous}. Let $G=(V,E)$ be a connected graph, and consider two species ($A$ and $B$). Each node may be in one of four states: vacant, occupied by an individual of species $A$, occupied by an individual of species $B$, or occupied simultaneously by one individual of each species, denoted by $0$, $A$, $B$, and $AB$, respectively. An individual located at a node of degree $k_i$ creates offspring in neighboring vacant nodes at rate $\lambda/k_i$. A node in state $A$ or $B$ becomes vacant at rate $\mu$, while a node in state $AB$ becomes vacant at a reduced symbiotic rate $\mu_s<\mu$.

A homogeneous mean-field approximation of the process with $\mu=1$ is given by
\begin{align}
    \dot{\rho_0} &= \rho_A + \rho_B - \lambda \rho_0\rho_T,\\
    \dot{\rho_A} &= -\rho_A + \rho_{AB} \mu_s + \lambda \rho_0 \left(\rho_{AB} + \rho_{A}\right) - \lambda \rho_A \left(\rho_{AB} + \rho_{B}\right),\\
    \dot{\rho_B} &= -\rho_B + \rho_{AB} \mu_s + \lambda \rho_0 \left(\rho_{AB} + \rho_{B}\right) - \lambda \rho_B \left(\rho_{AB} + \rho_{A}\right),\\
    \dot{\rho_{AB}} &= -2\mu_s \rho_{AB} + 2\lambda \rho_A \rho_B + \lambda \rho_{AB} \left(\rho_A + \rho_B\right),
\end{align}
where $\rho_X$, $X\in\{0,A,B,AB\}$, denotes the probability that a given node is in state $X$, and $\rho_T=\rho_A+\rho_B+2\rho_{AB}$ is the total prevalence.

\subsection{Symbiotic SIR}

Motivated by the structure of the 2SCP, we propose a variation of the classical SIR epidemic process on networks in which two diseases spread concurrently. Each individual is characterized by a state $(X,Y)\in\{s,i,r\}^2$, where the first (second) coordinate corresponds to the state with respect to disease $A$ (disease $B$). Transmission rates $\lambda_1$ and $\lambda_2$ are associated with diseases $A$ and $B$, respectively, and both diseases share a common recovery rate $\mu$. In analogy with the symbiotic mechanism of the 2SCP, we further assume that individuals co-infected with both diseases recover at a reduced rate $\bar{\mu}<\mu$.

The possible infection and recovery transitions are summarized by
\begin{align*}
    (s,y) \leftrightsquigarrow (i,y') &\overset{\lambda_1}{\longrightarrow} (i,y),\\
    (x,s) \leftrightsquigarrow (x', i)&\overset{\lambda_2}{\longrightarrow} (x, i),\\
    (i,y \neq i) &\overset{\mu}{\longrightarrow} (r,y),\\
    (x \neq i,i) &\overset{\mu}{\longrightarrow} (x,r),\\
    (i,i) &\overset{\bar{\mu}}{\longrightarrow} (r,r),
\end{align*}
where $x,x',y,y'\in\{s,i,r\}$ and the infectious contacts are denoted by $\leftrightsquigarrow$.

Although our model is minimal, it can be motivated by common scenarios where two diseases co-exist. For instance, the concurrent circulation of SARS-CoV-2 (COVID-19) and seasonal influenza, for which public health agencies explicitly discuss clinical management and testing in the presence of co-circulation and potential co-infection \citep{cdc_flucovid_testing, cdc_mmwr_coinfection_peds}. Empirically, co-infections of COVID with influenza have been documented in surveillance and clinical datasets, and meta-analyses report nonzero prevalence and potentially worse clinical outcomes in co-infected patients compared to mono-infected ones \citep{yan2023_coinfection_meta,golpour2025_coinfection_meta}. These observations motivate modeling frameworks in which the ``co-infected'' state is explicitly represented and may differ in recovery/clinical course from single-pathogen infection (here captured by $\bar\mu\neq\mu$), enabling studies of how co-infection can reshape peak burden, overlap effects, and epidemic duration.

\subsection{Literature review}
The study of interacting or coexisting epidemic processes has a substantial literature. For instance, \citet{kabir2019analysis} study an SIR process where an infection and awareness spread simultaneously and have an effect on each other's dynamics. This model has six possible states $(X,Y) \in \{S, I, R\} \times \{U, A\}$, where $U$ and $A$ refer to \textit{unaware} and \textit{aware}, respectively. These additional states represent a single absorbing transition, as it is not possible to go from aware back to unaware. 

Interacting epidemic dynamics have also been studied in settings where infection with one disease modifies susceptibility to another. \citet{funk2010interacting} study interacting epidemics on a single network, but where infection in disease $A$ grants a degree of immunity to disease $B$. 
Co-occurring epidemics have also been studied under the umbrella of competing epidemics \citep{darabi2014competitive, doshi2021competing, karrer2011competing, liu2019analysis, yang2017bi}. \citet{karrer2011competing} considers an SIR model where infection in either of two diseases confers future immunity to both. \citet{doshi2021competing} studies an SIS model of two diseases, giving conditions on the parameter space for global convergence to one of three possible states. \citet{darabi2014competitive} study competing SIS epidemics on multilayer networks. \citet{newman2013interacting} study two interacting SIR epidemics where infection with the second disease is only possible after being infected with the first one. 

To the best of our knowledge, the model studied here differs from existing approaches in that it considers two concurrent SIR epidemics with an explicitly represented co-infected state, distinct recovery rates for single and dual infection, and no assumed hierarchy, cross-immunity, or infection ordering between the two diseases, within a network mean-field framework.

\section{Model and basic properties}
We give the following equations for this Symbiotic SIR process in a graph with average degree $\langle k\rangle$.

\begin{align}
   (s,s)' &= -\langle k \rangle \sum_{y} (s,s)(i,y) \lambda_1 -\langle k \rangle\sum_{x} (s,s)(x,i) \lambda_2\\
   (i,s)' &= -(i,s)\mu - \langle k \rangle\sum_x (i,s)(x,i)\lambda_2 + \langle k \rangle \sum_y (s,s)(i,y) \lambda_1 \\
   (r,s)' &= (i, s)\mu - \langle k\rangle\sum_x (r,s)(x,i) \lambda_2 \\
   (s,i)' &= -\langle k\rangle\sum_y (s,i)(i, y) \lambda_1 - (s,i)\mu + \langle k\rangle\sum_x (s,s)(x, i) \lambda_2 \\
   (i, i)' &= \langle k\rangle\sum_y (s,i)(i,y)\lambda_1 + \langle k \rangle \sum_x (i,s)(x,i)\lambda_2 - (i,i)\bar{\mu}\\
   (r, i)' &= \langle k\rangle \sum_x (r,s)(x, i) \lambda_2 - (r,i)\mu\\
   (s,r)' &= -\langle k\rangle \sum_y (s,r)(i,y)\lambda_1 + (s,i)\mu\\
   (i, r)' &= \langle k\rangle\sum_y (s,r)(i, y)\lambda_1 - (i,r)\mu\\
   (r,r)' &= (i,i) \bar{\mu} + (i,r) \mu + (r,i) \mu 
\end{align}

Since there is no coupling between the dynamics of infection, the process behaves as two independent SIR processes, with respective epidemic thresholds $\mathcal{R}_{0,A} = \frac{\lambda_1 \langle k\rangle}{\mu}, \mathcal{R}_{0,B} = \frac{\lambda_2 \langle k\rangle}{\mu}$, giving our process a threshold of $\mathcal{R}_0 = \max(\mathcal{R}_{0,A}, \mathcal{R}_{0,B})$. Explicitly, by normalizing the population so that 
\[
\sum_{x,y\in\{s,i,r\}} (x,y)(t) = 1,
\]
the disease--free equilibrium (DFE) is
\[
E_0 = \big((s,s),(i,s),(r,s),(s,i),(i,i),(r,i),(s,r),(i,r),(r,r)\big)
= (1,0,0,0,0,0,0,0,0).
\]
At the DFE, co-infected states arise only through products of infected variables and do not contribute to the linearization. We proceed to obtain the reproductive number $\mathcal{R}_0$ via the next-generation method \citep{vandendriessche2002}. Letting 
\[
x =
\begin{pmatrix}
(i,s)\\
(s,i)
\end{pmatrix},
\]

the subsystem for $x$ can be written in the form
\[
x' = \mathcal{F}(x) - \mathcal{V}(x),
\]
where $\mathcal{F}$ contains the rates of appearance of new infections and $\mathcal{V}$ contains all other transitions. From the model equations,
\begin{align*}
(i,s)' &= \langle k\rangle \sum_{y}(s,s)(i,y)\lambda_1 - \mu(i,s) + \text{(higher--order terms)},\\
(s,i)' &= \langle k\rangle \sum_{x}(s,s)(x,i)\lambda_2 - \mu(s,i) + \text{(higher--order terms)}.
\end{align*}

Linearizing at the DFE, where $(s,s)=1$ and all infected compartments vanish, we obtain
\[
\mathcal{F}(x) =
\begin{pmatrix}
\langle k\rangle\lambda_1 (i,s)\\
\langle k\rangle\lambda_2 (s,i)
\end{pmatrix},
\qquad
\mathcal{V}(x) =
\begin{pmatrix}
\mu (i,s)\\
\mu (s,i)
\end{pmatrix}.
\]

The Jacobians of $\mathcal{F}$ and $\mathcal{V}$ at the DFE are therefore
\[
F = D\mathcal{F}(E_0) =
\begin{pmatrix}
\langle k\rangle\lambda_1 & 0\\
0 & \langle k\rangle\lambda_2
\end{pmatrix},
\qquad
V = D\mathcal{V}(E_0) =
\begin{pmatrix}
\mu & 0\\
0 & \mu
\end{pmatrix}.
\]
It follows that the next generation matrix is
\[
K = FV^{-1} = \begin{pmatrix}
\dfrac{\langle k\rangle\lambda_1}{\mu} & 0\\[6pt]
0 & \dfrac{\langle k\rangle\lambda_2}{\mu}
\end{pmatrix},
\]
and hence the basic reproduction number of the coupled system is the spectral radius
\[
\mathcal R_0 = \rho(K) = \max(\mathcal R_{0,A},\,\mathcal R_{0,B}).
\]
The key difference that this model brings comes from the `symbiotic' recovery. To focus on this, suppose one has $\lambda_1 = \lambda_2 = \lambda$ (i.e. equal transmission rates) and let us study the regime when $\bar{\mu} < \mu$. When $\lambda_1=\lambda_2=\lambda$, the system is invariant under exchange of diseases $A$ and $B$. More precisely, the transformation
\[
(x,y)\longmapsto (y,x), \qquad x,y\in\{s,i,r\},
\]
maps solutions of the system to solutions. Consequently, if the initial conditions satisfy
\[
(i,s)=(s,i), \quad (r,s)=(s,r), \quad (r,i)=(i,r),
\]
then these equalities hold for all $t\geq 0$ by uniqueness of solutions. Under this symmetry, the dynamics evolve on the invariant subspace defined by
\[
a=(i,s)=(s,i),\quad b=(r,s)=(s,r),\quad d=(r,i)=(i,r),
\]
with $u = (s,s)$, $c = (i,i)$, and $v = (r,r)$ unchanged. This allows the original nine dimensional system to be reduced to a six dimensional system on this invariant subspace. In what follows, we will restrict ourselves to this symmetric case.

\subsection{Co-infection burden and monotonicity in $\bar{\mu}$}
Defining the total infection mass $M(t) = a(t) + c(t) + d(t)$, one gets 
\begin{equation}\label{eq:symmetric-coinfect}
    c'(t) = 2 \langle k\rangle \lambda a(t) M(t) - \bar{\mu} c(t)
\end{equation}
One can see that decreasing $\bar{\mu}$ increases the prevalence of co-infectious individuals, namely, $c(t)$ is decreasing in $\bar{\mu}$. In fact, if for fixed $\bar{\mu}$ one defines the cumulative co-infection burden $B(\bar{\mu}) = \int_0^\infty c(t) \, dt$, this will also be decreasing in $\bar{\mu}$. A numerical illustration of this can be found in Figure \ref{fig:coinfection_side_by_side}. Formally, one can state the following result:

\begin{figure}[t]
    \centering
    \begin{subfigure}{0.48\linewidth}
        \centering
        \includegraphics[width=\linewidth]{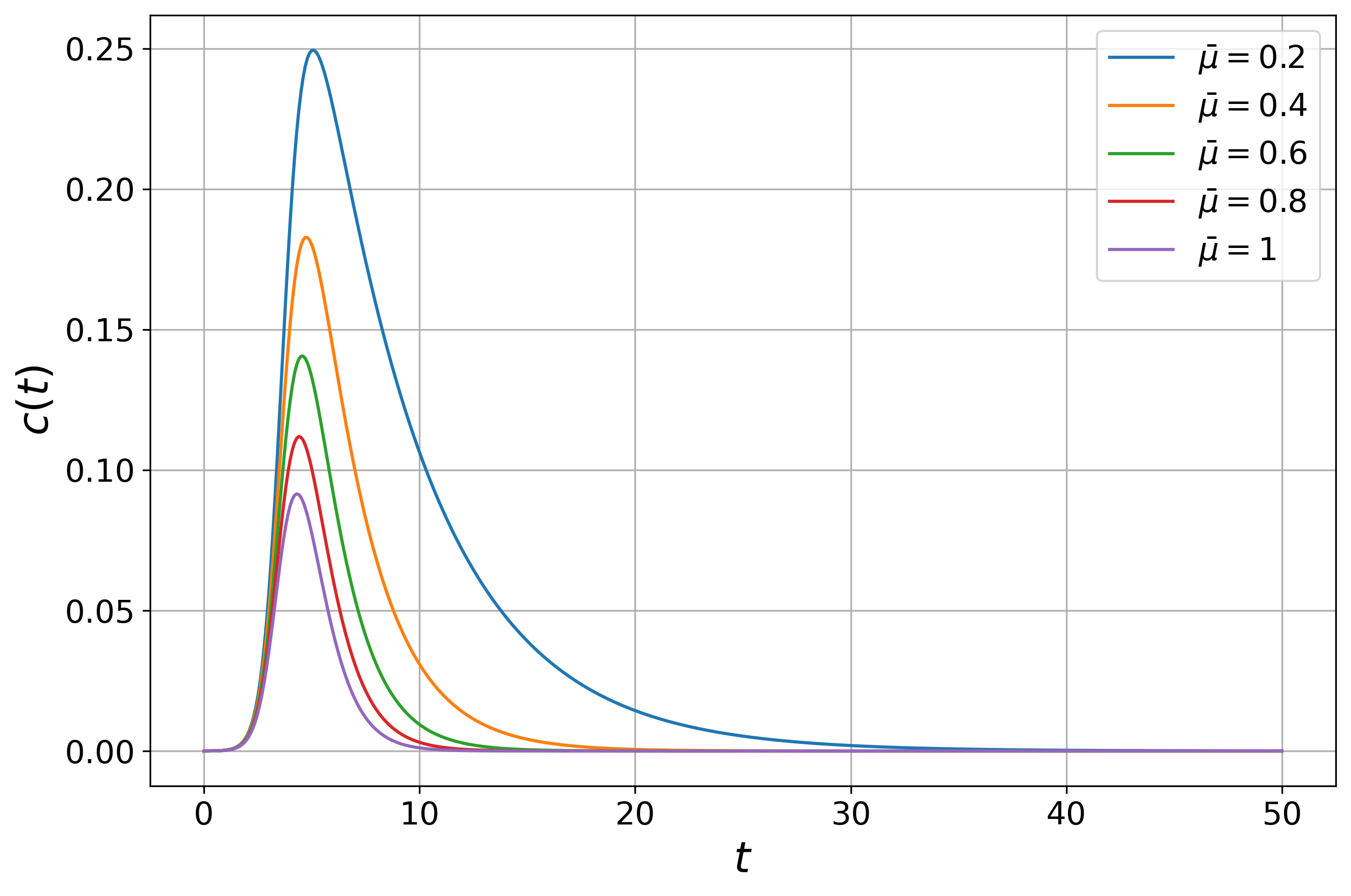}
        \caption{Co-infection $c(t)$ decreasing in $\bar{\mu}$.}
        \label{fig:c_monotone}
    \end{subfigure}
    \hfill
    \begin{subfigure}{0.48\linewidth}
        \centering
        \includegraphics[width=\linewidth]{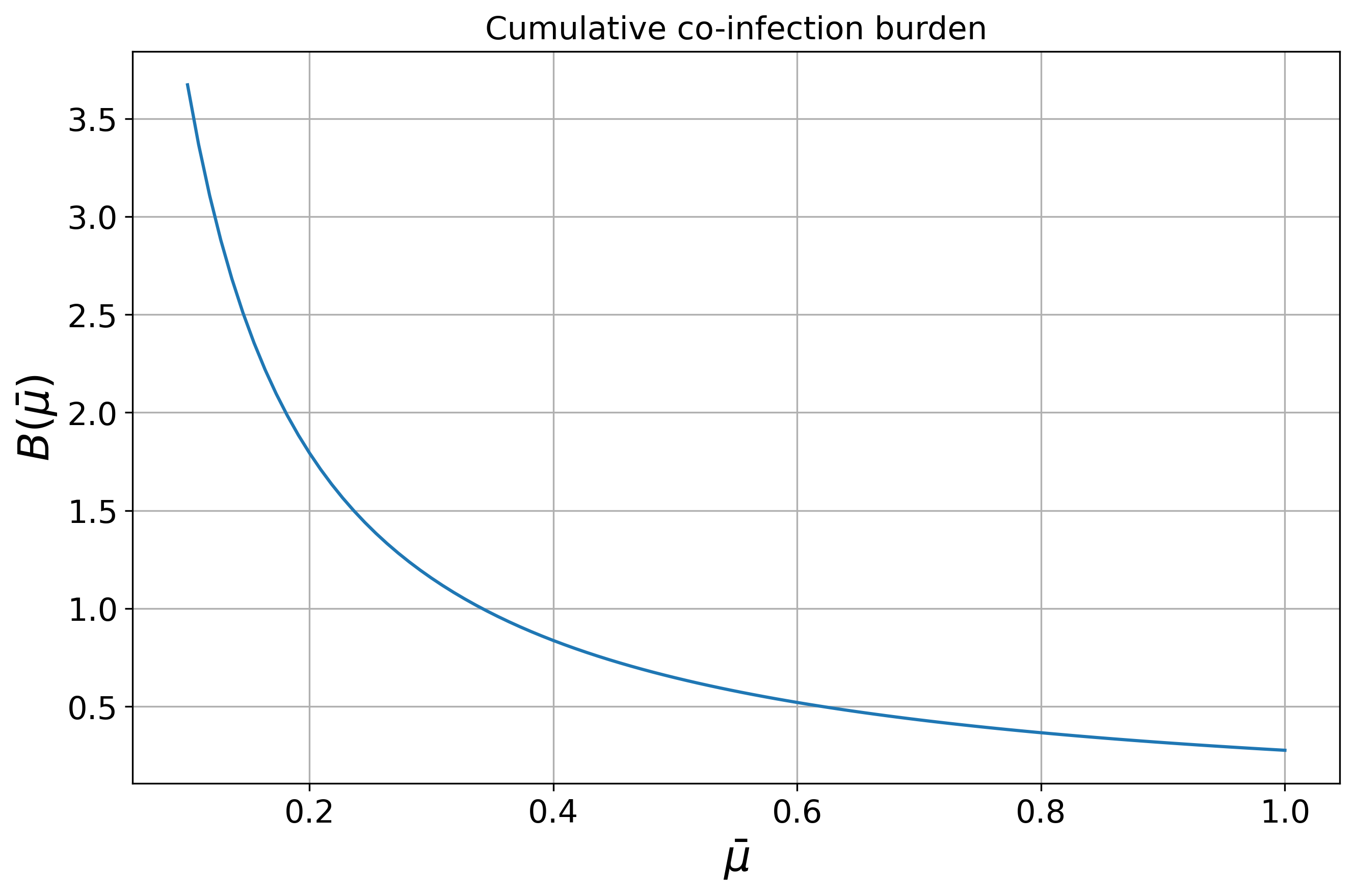}
        \caption{Cumulative co-infection burden $B(\bar{\mu})$.}
        \label{fig:cum_burden}
    \end{subfigure}
    \caption{Coinfection dynamics for different values of $\bar{\mu}$, with $\langle k\rangle=10, \lambda = 0.25, \mu = 1$.}
    \label{fig:coinfection_side_by_side}
\end{figure}

\begin{prop}
Let $g: [0, \infty) \longrightarrow [0, \infty)$ be a measurable function and $c_0 \geq 0$. For $\bar{\mu} > 0$, let $c_{\bar{\mu}}(t)$ solve 
\[
c^{'}_{\bar{\mu}}(t) = g(t) - \bar{\mu}c_{\bar{\mu}}(t).
\]
If $0 < \bar{\mu}_1 < \bar{\mu}_2$, then 
\[
\forall t \geq 0: \quad c_{\bar{\mu}_1}(t) > c_{\bar{\mu}_2}(t),
\]
and
\[
\forall T > 0:\quad \int_{0}^{T} c_{\bar{\mu}_1}(t) \, dt \geq \int_{0}^{T} c_{\bar{\mu}_2}(t) \, dt.
\]
\end{prop}
\begin{proof}
    See from Equation \ref{eq:symmetric-coinfect} that $c'$ has the form required by this proposition. Then, by variation of constants,
    \[
    c_{\bar{\mu}}(t) = e^{-\bar{\mu}t} c_0 + \int_0^t e^{-\bar{\mu} (t-s)} g(s) \, ds,
    \]
    which is readily seen to be decreasing in $\bar{\mu}$. The integral inequality follows by monotonicity of the integral.
\end{proof}

\subsection{Epidemic duration}
Define $i_A = \sum_y (i, y), i_B = \sum_x (x,i)$ (the total infectious individuals of disease $A$ and $B$ respectively). The total number of infectious individuals is $J = i_A + i_B - (i,i)$. Under our symmetry assumption, this reduces to $J(t) = 2a(t) + 2d(t) + c(t)$. Also following from our symmetry is a lower bound on the epidemic duration which \textit{increases} when $\bar{\mu}$ decreases. Indeed, from Equation \ref{eq:symmetric-coinfect} one sees that 
\begin{equation}
    c'(t) \geq -\bar{\mu} c(t).
\end{equation}
Consider a time $t_0$ such that $c(t_0) > \varepsilon$ for some threshold $\varepsilon > 0$. Integrating, one gets that for all $t \geq t_0$, 
\begin{equation}
    c(t) \geq c(t_0) e^{-\bar{\mu}(t-t_0)}.
\end{equation}
In this regime, one can define the co-infection duration as $T_c (\varepsilon) = \inf\{t : c(t) < \varepsilon\}$. Given that $c(t_0) > \varepsilon$, then it readily follows that $c(t) \geq \varepsilon$ when $t \leq t_0 + \frac{1}{\bar{\mu}} \ln\left(\frac{c(t_0)}{\varepsilon}\right)$, so $T_c (\varepsilon) \geq t_0 + \frac{1}{\bar{\mu}} \ln\left(\frac{c(t_0)}{\varepsilon}\right)$. Since $J(t) \geq c(t)$, the epidemic duration which can be defined as
\begin{equation}
    T_J (\varepsilon) = \inf\{t : J(t) < \varepsilon\},
\end{equation}
must satisfy 
\begin{equation}
    T_J (\varepsilon) \geq T_c (\varepsilon) \geq t_0 + \frac{1}{\bar{\mu}} \ln\left(\frac{c(t_0)}{\varepsilon}\right).
\end{equation}
It is clear that decreasing $\bar{\mu}$ will increase this bound.

\begin{figure}
    \centering
    \includegraphics[width=0.5\linewidth]{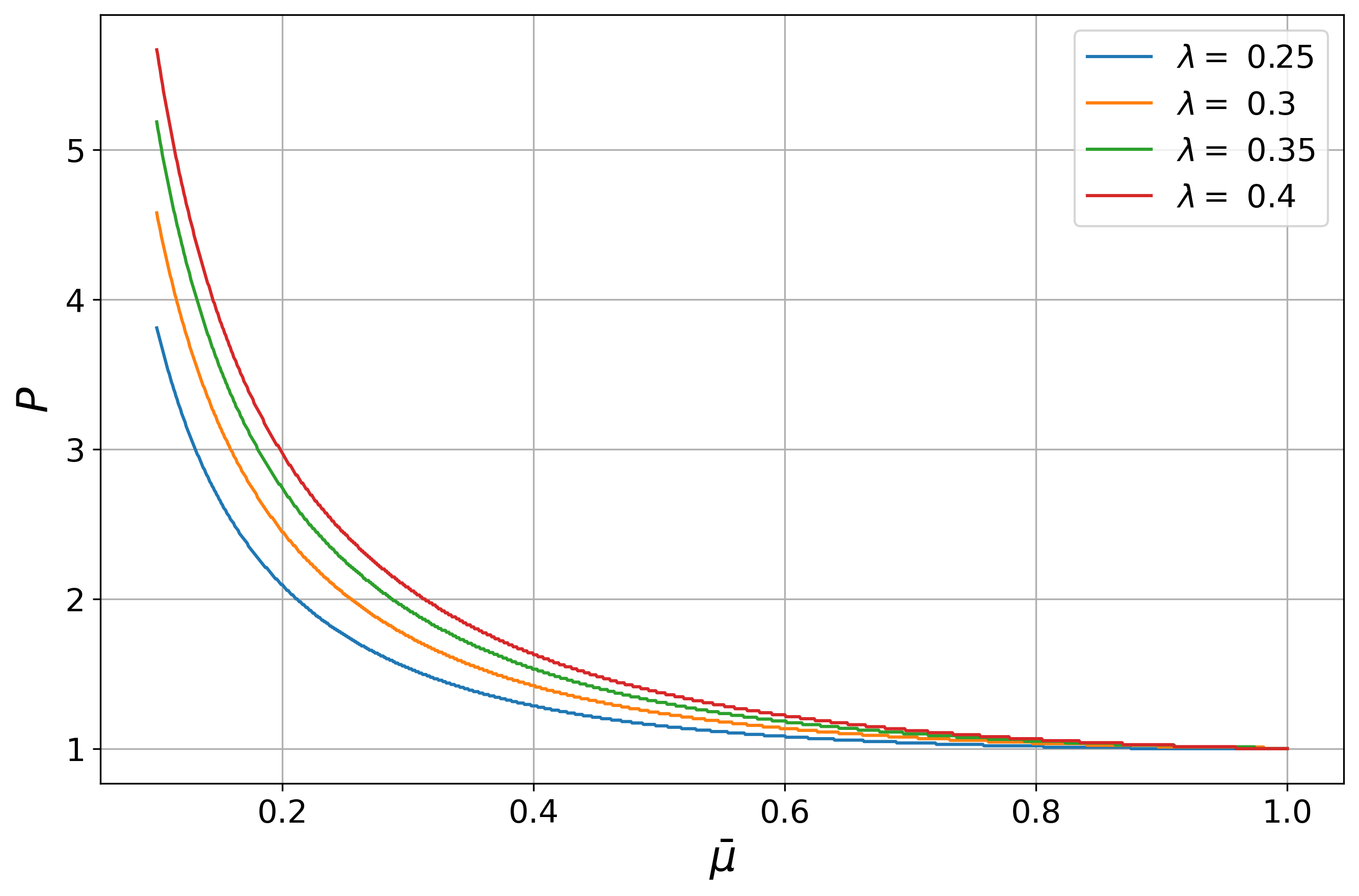}
    \caption{Ratio of duration of the epidemic with $\bar{\mu} < \mu = 1$ for different values of $\lambda$} 
    \label{fig:ratio_duration}
\end{figure}

Numerically, one can see that when co-infectious individuals recover at a slower rate (i.e. $\bar{\mu} < \mu$) a longer epidemic-duration follows.  
Fix $\langle k \rangle = 10, \mu = 1$ and consider $\mathcal{D}(x)$ the duration of the epidemic when $\bar{\mu} = x$. A quantity of interest is the ratio $P = \frac{\mathcal{D}(x)}{\mathcal{D}(1)}$. As we see in Figure \ref{fig:ratio_duration}, the smaller $\bar{\mu}$ is with respect to $\mu$, the larger the duration of the epidemic with respect to the case $\mu = \bar{\mu}$. For example, if $\lambda = 0.25$ and $\bar{\mu} = 0.2$, the epidemic has a duration of approximately twice as long as when $\bar{\mu} = \mu = 1$.

\subsection{Peak prevalence and sensitivity analysis}
\begin{figure}
    \centering
    \includegraphics[width=0.5\linewidth]{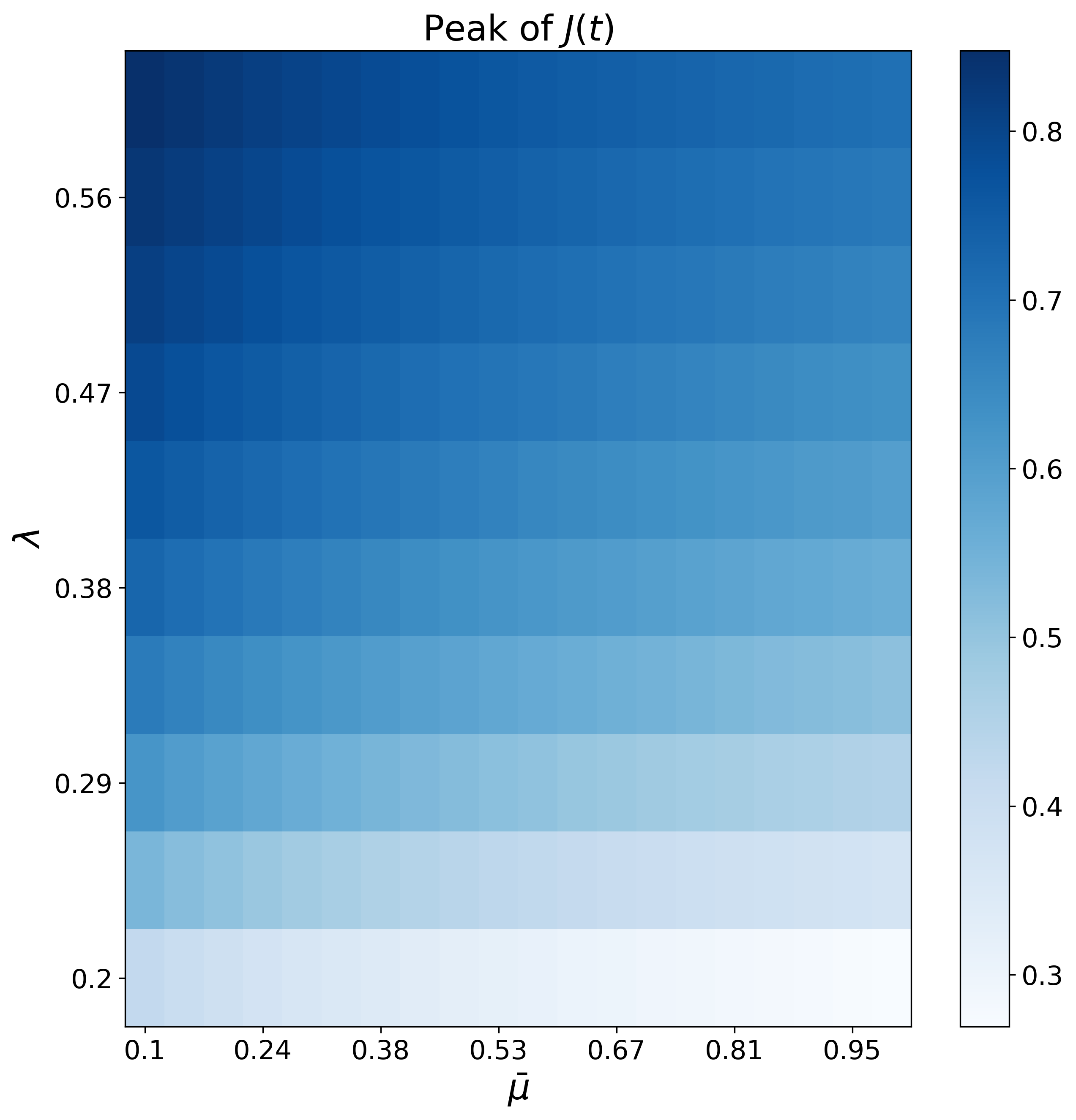}
    \caption{Maximum value of $J(t)$ for $\lambda = 0.2, 0.24, \dots, 0.6$ and $\bar{\mu} = 0.1, \dots,1$, with $\langle k\rangle = 10$ and $\mu = 1$.}
    \label{fig:peak_J}
\end{figure}

The longer period of recovery for those co-infected with both diseases seems to have an impact in the peak of $J(t)$, i.e. its maximum value. In a high transmission regime, fixing the same parameters as before but now varying $\lambda$ in the range $[0.2, 0.6]$, we observe the peak is larger the smaller $\bar{\mu}$ is. See Figure \ref{fig:peak_J}.

\begin{figure}
    \centering
    \includegraphics[width=0.5\linewidth]{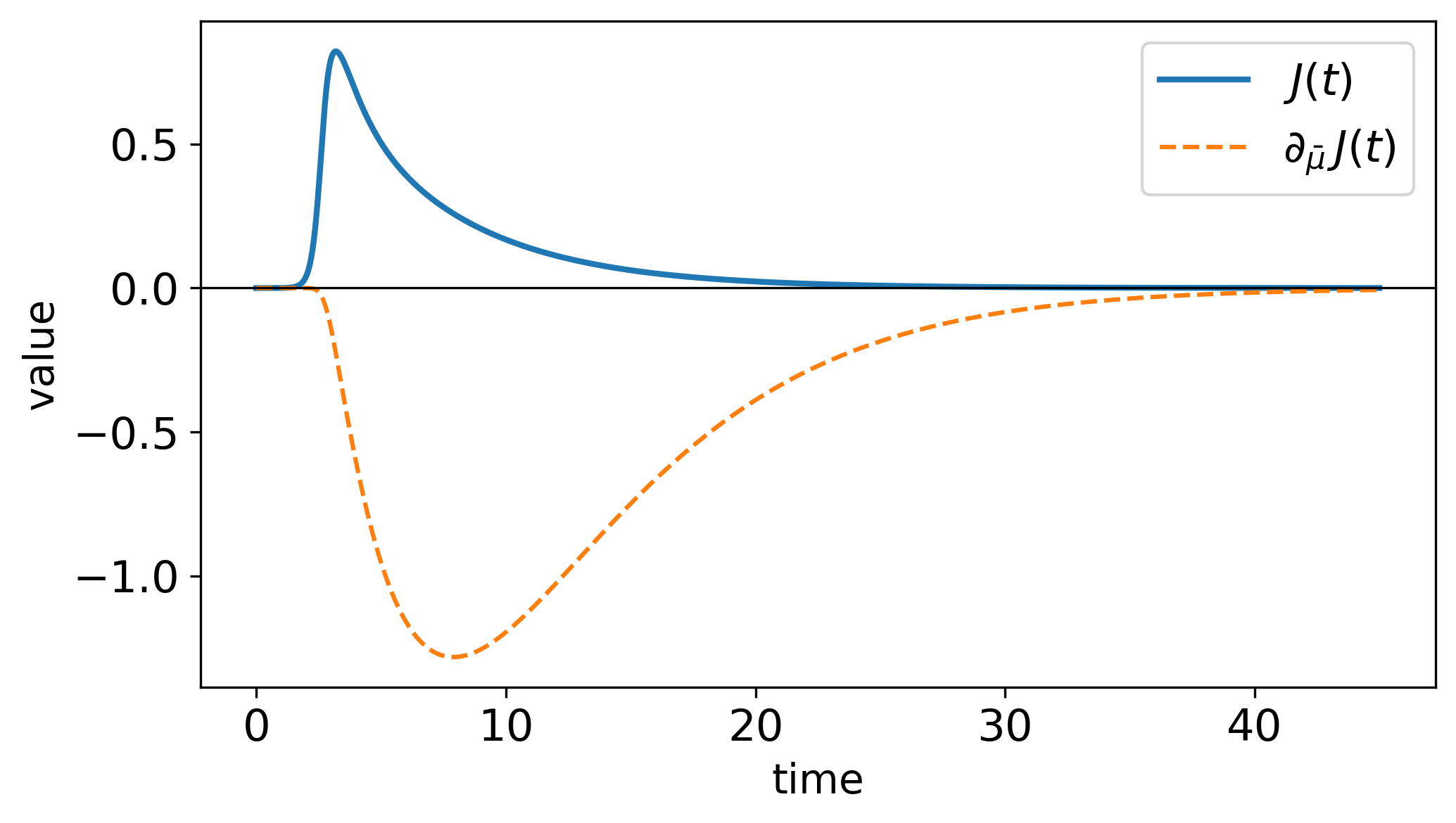}
    \caption{Sensitivity $\partial_{\bar\mu}J(t)$ and $J(t)$ in a numerical solution for the case $\langle k \rangle = 10, \lambda_1 = \lambda_2 = 0.3, \mu = 1, \bar{\mu} = 0.5$}
    \label{fig:djT_sensitivity}
\end{figure}

This can be partially explained as follows. Differentiating $J$ one gets 
\begin{equation}
    J'(t)
=
2\langle k\rangle\lambda\,(u(t)+b(t))\,M(t)
-
\big(2\mu\,a(t)+2\mu\,d(t)+\bar\mu\,c(t)\big).
\end{equation}
When $\bar{\mu} c(t)$ is smaller, everything else equal, $J'$ is larger when $c(t) > 0$. Further evidence of the aforementioned can be obtained by a sensitivity analysis. Let $Y' = f(Y,\bar\mu)$, with $Y=(u,a,b,c,d,v)^\top$, and define the sensitivity
$\Sigma(t)=\partial_{\bar\mu}Y(t) = (\sigma_u, \sigma_a, \dots, \sigma_v)^\top$. Then $\Sigma$ satisfies
\[
\Sigma'(t)=D_Y f(Y(t),\bar\mu)\,\Sigma(t)+\partial_{\bar\mu}f(Y(t),\bar\mu),
\qquad \Sigma(0)=0,
\]
where, for the symmetric reduced system,
\[
\partial_{\bar\mu}f(Y,\bar\mu)=(0,0,0,-c,0,c)^\top.
\]
Moreover, with $J(t)=2a(t)+c(t)+2d(t)$,
\[
\partial_{\bar\mu}J(t)=2 \sigma_a(t)+ \sigma_c (t) + 2\sigma_d (t).
\]
One can numerically solve for $\partial_{\bar\mu}J$ and see how the sensitivity is negative towards the peak, implying that slower co-infectious recovery raises the epidemic peak to first order. See Figure \ref{fig:djT_sensitivity}.

\section{Conclusion}

We have introduced and analyzed a symmetric two-disease SIR co-infection model on networks, motivated by the symbiotic two-species contact process and by real scenarios in which multiple pathogens co-circulate. The model explicitly represents co-infected individuals and allows for a distinct recovery rate in the co-infected state, while retaining absorbing recovered states for each disease. Within a mean-field network approximation, we derived the basic reproduction number of the coupled system and showed that, at the disease-free equilibrium, invasion thresholds reduce to those of two independent SIR processes.

Focusing on the symmetric transmission regime $\lambda_1=\lambda_2$, we exploited an exchange symmetry between the two diseases to reduce the original nine-dimensional system to a six-dimensional invariant subsystem. This reduction enabled analytic insight into the role of the co-infection recovery rate $\bar{\mu}$. In particular, we proved that decreasing $\bar{\mu}$ monotonically increases both the instantaneous prevalence of co-infection and the cumulative co-infection burden over finite time horizons. Using a Gr\"onwall-type argument, we further established a lower bound on epidemic duration that scales inversely with $\bar{\mu}$, showing that slower recovery of co-infected individuals necessarily prolongs the epidemic tail.

Numerical simulations complemented these results and revealed an additional effect of reduced co-infection recovery: in a sufficiently high-transmission regime, smaller values of $\bar{\mu}$ are associated not only with longer epidemics but also with a larger peak in the total infectious prevalence. This observation was supported by a sensitivity-equation analysis, which showed that the sensitivity of the total infection burden with respect to $\bar{\mu}$ is negative near the epidemic peak, indicating that slower co-infection recovery raises the peak to first order.

The present work does not aim to capture pathogen-specific biological details. Nevertheless, it highlights how explicitly modeling co-infection and altered recovery can qualitatively affect epidemic duration and peak burden, even in the absence of cross-immunity, hierarchical infection ordering, or endemic equilibria. Possible extensions include incorporating higher order contact structures, further exploration of the asymmetric transmission rates regime, or stochastic effects, as well as exploring how vaccination or partial immunity might interact with co-infection dynamics. More broadly, the results emphasize that co-infection-specific parameters can play a significant role in shaping transient epidemic dynamics, beyond what is predicted by single-disease models.

\bibliographystyle{elsarticle-num-names} 
\bibliography{ref}

\end{document}